\documentclass[11pt]{article} 

\usepackage[utf8]{inputenc} 

\usepackage{geometry} 
\usepackage{graphicx} 

\usepackage{booktabs} 
\usepackage{array} 
\usepackage{paralist} 
\usepackage{verbatim} 
\usepackage{subfig} 
\usepackage{color}
\usepackage{fancyhdr} 
\usepackage{sectsty}
\allsectionsfont{\sffamily\mdseries\upshape} 

\usepackage[nottoc,notlof,notlot]{tocbibind} 
\usepackage[titles,subfigure]{tocloft} 

\newcommand{\beq}{\begin{equation}}
\newcommand{\eeq}{\end{equation}}
\newcommand{\de}{\mbox{d}}

\newcommand{\bds}{\begin{displaystyle}}
\newcommand{\eds}{\end{displaystyle}}
\newcommand{\ie}{{\em i.e.\ }}

\newcommand{\red}{\color{black}}

\title{Single-molecule modeling of mRNA degradation by miRNA: Lessons from data}
\author{Celine Sin, Davide Chiarugi, Angelo Valleriani\footnote{email: {\tt angelo.valleriani@mpikg.mpg.de}}\\
Department of Theory and Bio-Systems\\
Max Planck Institute of Colloids and Interfaces\\
14424 Potsdam, Germany}
\date{} 

\begin{document}
\maketitle
\begin{abstract} 
Recent experimental results on the effect of miRNA on the decay of its target mRNA have been analyzed against a previously hypothesized single molecule degradation pathway. According to that hypothesis, the silencing complex (miRISC) first interacts with its target mRNA and then recruits the protein complexes associated with NOT1 and PAN3 to trigger deadenylation (and subsequent degradation) of the target mRNA. Our analysis of the experimental decay patterns allowed us to refine the structure of the degradation pathways at the single molecule level.  Surprisingly, we found that if the previously hypothesized network was correct, only about 7\% of the target mRNA would be regulated by the miRNA mechanism, which is inconsistent with the available knowledge. 
Based on systematic data analysis, we propose the alternative hypothesis that NOT1 interacts with miRISC before binding to the target mRNA. Moreover, we show that when miRISC binds alone to the target mRNA, the mRNA is degraded more slowly, probably through a deadenylation-independent pathway.
The new biochemical pathway we propose both fits the data and paves the way for new experimental work to identify new interactions.

\end{abstract}


\section{Introduction}
\subsection{Background}
In living cells, the level of protein expression is thoroughly regulated. Many crucial processes for this regulation occur at the post-transcriptional level. In this context, the control mechanisms acting on  messenger RNAs (mRNAs) play a pivotal role. Indeed, living cells are endowed with a number of biochemical pathways converging on cytosolic mRNAs, that serve to enhance or repress gene expression. These pathways are known to operate (1) either by enhancement \cite{Orom2008a,Vasudevan2007a} or repression \cite{Huntzinger2011a,Izaurralde2012a} of translation or (2) modulating mRNA lifetimes \cite{Huntzinger2011a,Repetto2012a,Izaurralde2012a,Cairrao2009a}.
The global picture emerging from the growing body of experimental evidence, depicts a complex interaction network which affects the mRNAs available for translation. This network is composed of several biochemical pathways,  often interwoven and cross-talking \cite{Ma2010a,Jing2005a}, and involves mRNA binding proteins as well as non coding RNAs \cite{Houseley2009a,Tay2014a,Kartha2014a}.
While there are a number of mechanisms responsible for mRNA degradation in eukaryotic cells \cite{Houseley2009a}, the decay of messages mediated by micro-RNAs (miRNAs) plays a prominent role in the control of gene expression \cite{Huntzinger2011a, Belasco2010a, Lu2012a}.

Despite extensive study, the topology and dynamics of the miRNA-mediated mRNA degradation pathway are still unclear. One of the main challenges stems from the fact that intermediate states of the pathway are unknown or difficult to quantify; experimentally, it is only feasible to measure the final state of the pathway (e.g. the decay pattern of the target mRNA). Bridging the gap between observed {\em decay patterns} and {\em degradation pathways} is non-trivial \cite{Deneke2013a}, since the former refer to a population average and the latter refers to the single-molecule stochastic process of degradation. Here we apply a rigorous strategy to reconstruct the miRNA-mediated degradation pathway, starting from experimentally measured decay patterns. Surprisingly, we find the previously proposed pathway not consistent with the experimental data. We propose an alternative model which both fits the decay pattern and allows us to gain some insight on the network topology.

\subsection{The experimental data}

\begin{figure}[h!]
\begin{center}
\includegraphics[scale = 0.4]{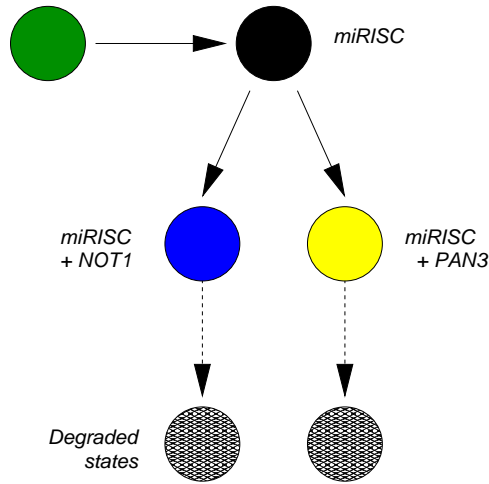} 
\end{center}
 \caption{The biochemical degradation network hypothesized by Braun {\em et al.} \cite{Braun2011a} for the degradation of a target mRNA by miRNA in {\em D.\ melanogaster} S2 cells. According to this network, the mRNA in its initial state (green circle) first binds to the miRISC complex thus leading to a new biochemical state (black circle). The miRISC and its target then recruit the proteins NOT1 and/or PAN3, leading to the two states indicated with the blue and yellow circle, respectively. From these two states the mRNA is finally degraded through a complex sequence of events including deadenylation followed by decapping \cite{Huntzinger2011a}. This unspecified sequence of events is indicated with dotted arrows in all figures of this paper.}
\label{fighyp}
\end{figure}

\begin{figure}[h!]
\begin{center}
\includegraphics[scale = 0.6]{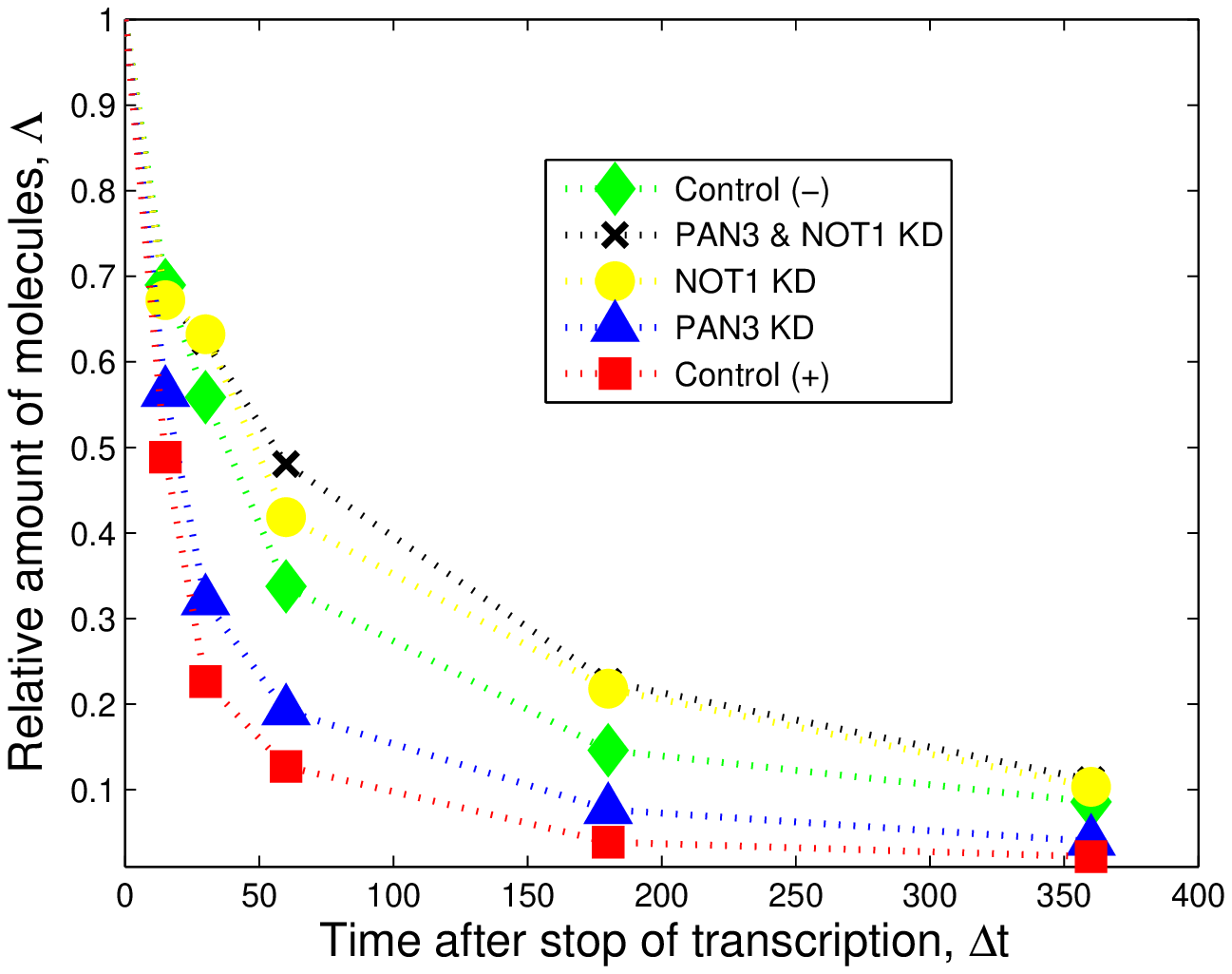} 
\end{center}
 \caption{The experiment performed by Braun {\em et al.}\cite{Braun2011a} consists of knocking down several permutations of the target's degradation factors. The ``Control (-)" data result from an experimental set-up in which the miR9-b is not expressed; the ``PAN3 \& NOT1 KD" data result from an experimental set-up in which only the miR9-b is expressed but both NOT1 and PAN3 are knocked down; the ``NOT1 KD" data result from a set-up in which miR9-b and PAN3 are expressed while NOT1 is knocked down; the ``PAN3 KD" data result from a set-up in which miR9-b and NOT1 are expressed while PAN3 is knocked-down; the ``Control (+)" data result from a set-up in which all three factors are expressed. The data have been extracted from figure 7A of \cite{Braun2011a}. The numerical values are reported in the supplementary materials.}
\label{figdata}
\end{figure}

To clarify the interplay of the various factors in a degradation pathway involving miRNA, and the protein complexes NOT1 and PAN3, Braun {\em et al.} \cite{Braun2011a} performed a series of controlled knock-down experiments in {\em D.\ melanogaster} S2 cells containing constructs for the miRNA miR-9b, its target mRNAs, namely the F-Luc-Nerfin mRNAs, and the factors NOT1 and PAN3, which are known to trigger mRNA deadenylation. In each experiment, a subset of NOT1, PAN3 and/or the miRNA miR-9b were selectively knocked down, yielding cell lines expressing different combinations of those factors: a control line without the miRNA, a cell line with miR-9b only, a cell line with NOT1+miR-9b (but not PAN3), a cell line with PAN3+miR-9b (but not NOT1),  and finally a cell line with all three factors NOT1+PAN3+miR9-b.  After steady state expression of the factors, the transcription of mRNA was blocked and the decay patterns over time {\red for three independent biological replica} were measured. {\red The average decay pattern from these three replicas was reported in \cite{Braun2011a} from which we could extract the data shown in figure \ref{figdata}}.

The conclusion of this detailed experimental study is that NOT1 is a more relevant factor than PAN3 in destabilizing the mRNA \cite{Braun2011a}. When only NOT1 is knocked down, the decay of F-Luc-Nerfin mRNA is significantly slower (yellow curve, figure \ref{figdata}) than the control (red curve, figure \ref{figdata}). In contrast, the effect of PAN3 knock down is less significant (blue curve, figure \ref{figdata}). These findings apparently confirm that the degradation pathway through NOT1 in figure \ref{fighyp} is the most prominent pathway for degradation of the target mRNA.
Although this conclusion is relatively robust, the published analyses do not validate the hypothesized biochemical degradation pathway given in figure \ref{fighyp}. Indeed, in the negative control (green curve in figure \ref{figdata}), the miRNA is knocked-down so that the formation of a specific silencing complex miRISC is suppressed, yet the target mRNA still decays. Additionally, when only the miRNA is expressed while PAN3 and NOT1 are knocked-down, the target mRNA decays (black curve in figure \ref{figdata}), but is definitively more stable than in the negative control. Both of these cases suggest that the model hypothesized in figure \ref{fighyp} should be expanded to include additional degradation pathways.

An important conceptual consideration is that figure \ref{fighyp} depicts degradation from the single-molecule perspective whereas the curves in figure \ref{figdata} are averages as a function of time. Therefore, our strategy consists of starting with the network shown in figure \ref{fighyp} and validating it against the experimental decay patterns. At the same time, we will propose alternative parsimonious extensions of the network when the validation fails. {\red 
In particular, we will find that when the miRISC complex interacts with the mRNA alone, it seems to stabilize the mRNA and trigger a deadenylation independent degradation of the mRNA.  Furthermore, we will show that the data supports the hypothesis that miRISC binds to NOT1 {\em before} recruiting the target mRNA and that there is a strong enhancement of mRNA recruitment when PAN3 is also present.}  
 
\section{Methods}
As previously mentioned, the relationship between {\em degradation pathways} (such as the one in figure \ref{fighyp}) and {\em decay patterns} (such as those in figure \ref{figdata}) is not trivial. If the decay pattern was exponential, the halftime of the mRNA population estimated from the decay pattern would be directly related to the rate of decay of single molecules. {\red The analysis of the decays shown in figure 2 shows that a model based on a single exponential function results in a poor fit; more complex models are preferred even in light of evaluations based on the Akaike Information Criteria (AIC). Furthermore, if the decay curves could be well described by a model based on a single exponential, the traces would appear as straight lines when plotted in a linear-log scale}  (see figure 7A in \cite{Braun2011a}). 

\begin{figure}[h!]
\begin{center}
\includegraphics[scale = 0.4]{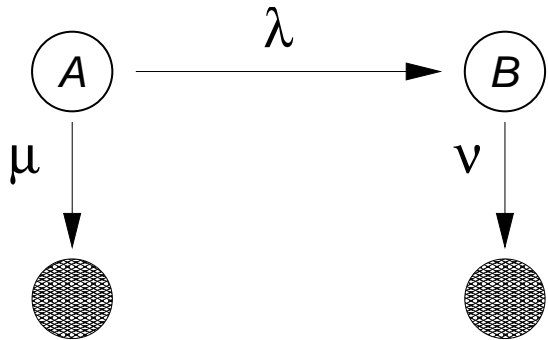} 
\end{center}
 \caption{This two-state model is able to fit all of the decay patterns in figure \ref{figdata}, by an appropriate choice of the three fitting parameters, $\lambda$, $\mu$, $\nu$. In this context, the mRNA is initially in the state $A$. From this state it can be degraded directly with a rate $\mu$ or change biochemical state, with rate $\lambda$, from which it is then degraded with rate $\nu$. Although this network performs a better fit of the data and is preferred to the exponential fitting based on the AIC criterion, the interpretation of the states is unclear. The fitting curves and the parameters are discussed in the supplementary materials.}
\label{figsimple}
\end{figure}

The issue of relating complex degradation pathways to decay patterns has been tackled in \cite{Deneke2013a} and in \cite{Wu2013a}. In \cite{Deneke2013a} it was shown that decay patterns similar to those depicted in figure \ref{figdata} can be generated by single-molecule networks satisfying certain properties, if one assumes that the transitions between biochemical states can be modeled as first-order chemical reactions. The mathematics supporting this reasoning was presented in \cite{Deneke2013a} and is summarized in the supplementary materials where we show how to derive the necessary mathematical functions using first passage time methods \cite{Valleriani2008a, Keller2012a, Valleriani2014a}. In order to generate decay patterns such as those in figure \ref{figdata}, the corresponding single-molecule degradation pathway must be composed of at least two states from which degradation is possible. Thus, in principle, one can either hypothesize a network of states that represents the biochemical pathway of degradation based on predictions and prior knowledge, or one can use the mathematical relationships mentioned above to find the most parsimonious network that fits that data. The most parsimonious network of states that is able to fit each one of the curves in figure \ref{figdata} is given by a two-state model depicted in figure \ref{figsimple}.

While the network in figure \ref{figsimple} results in a definitively better fit to the data and thus could be used to derive quantities such as the average lifetime and the age dependent degradation rate, it does not address the question as to whether or not the network in figure \ref{fighyp} is a suitable framework for the decay patterns observed in figure \ref{figdata}. To address this question we employ a hierarchical strategy: (i) we start by fitting the negative control decay pattern (the green trace ``Control (-)" in figure \ref{figdata}) to the most parsimonious model (figure \ref{figsimple}) and thereby fix the corresponding three rates; (ii) we then consider the next decay pattern with one additional decay factor active and enlarge the network of states to accommodate the additional decay factor.  We continue until each curve has been evaluated and the corresponding network is built.

{\red The details of  the functions used to perform the fit can be found in the Supplementary Materials. In particular, section S1.1 provides the general aspects of the mathematical background required for the purpose of this paper and section S1.2 gives the explicit formulas used for the fit. The parameters were estimated using nonlinear fit algorithms available in MATLAB (see section S1.4).}  

\section{Results}

Based on the hierarchical strategy above, we start with the ``Control (-)" curve (green decay pattern, figure \ref{figdata}). This curve describes the decay of the mRNA when none of the degradation factors (miRNA, NOT1 and PAN3) are present.  In the framework of the hypothesized network in figure \ref{fighyp}, this would correspond to all downstream processes inactive thus reducing the network to just the first state (green circle).  This is obviously not sufficient to explain the observed decay, since a single state without decay would produce a horizontal line (\ie no decay).  
This one-state scenario is also not consistent with biological reality. Indeed, even the most stable cellular macromolecule is eventually degraded. In particular, there are many biochemical pathways devoted to mRNA degradation \cite{Houseley2009a}.

\begin{figure}[h!]
\begin{center}
\includegraphics[scale = 0.6]{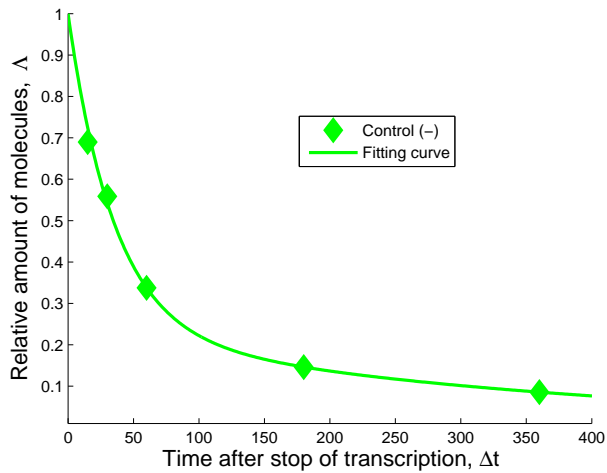} 
\end{center}
 \caption{The negative control decay pattern can be fitted with the single molecule network from figure \ref{figsimple}. This delivers the rates $\lambda$, $\mu$ and $\nu$ that will be kept constant through any successive enlargement of the network when considering the other decay patterns. The values of the rates are: $\lambda = 0.0008\, \mbox{min}^{-1}$, $\mu = 0.0276\, \mbox{min}^{-1}$, $\nu = 0.0028\, \mbox{min}^{-1}$, with confidence intervals $[0.0002, \, 0.0013]$, $[0.0229,\, 0.0324]$, $[0.0018,\, 0.0038]$, respectively.}
\label{figgreen}
\end{figure}

In the absence of further information, we fit the green ``Control (-)" curve of figure \ref{figsimple} to the most parsimonious (or minimal) network that still captures the dynamics of the data (figure \ref{figgreen}). {\red We can interpret the need for such network by saying that} in absence of miRNA-dependent degradation, mRNA molecules can be degraded through two pathways, differing by their kinetic features (see figure \ref{figsimple}). The first class of pathways (governed by the rate $\mu$ in figure \ref{figsimple}) is characterized by a single step and it can be representative of the set of ``constitutive'' reactions which target mRNAs non-specifically and are catalyzed by enzyme complexes such as the exosome \cite{Houseley2009a}. The second class of pathways (along the path $\lambda$ and $\nu$ in figure \ref{figsimple}) exhibits two steps and represents the degradation processes (independent of miRNAs) passing through a control step of a more complex degradation pathway (e.g. the preliminary binding of specific proteins to the target mRNA). One such example independent of miRNA and the NOT1/PAN3 factors is the ARE-mediated degradation pathway \cite{Cairrao2009a, Helfer2012a}. {\red We should stress that the network to fit the green ``Control(-)'' curve of figure \ref{figsimple} was not foreseen in the pathway proposed in \cite{Braun2011a} and that we introduce it in order to consistently perform a fit for each of the experimental decay patterns. The existence and the strength of such additional, constitutive degradation pathway may be dependent on the species of mRNA and on the growth conditions of the cell culture.}

\begin{figure}[h!]
\begin{center}
\includegraphics[scale = 0.4]{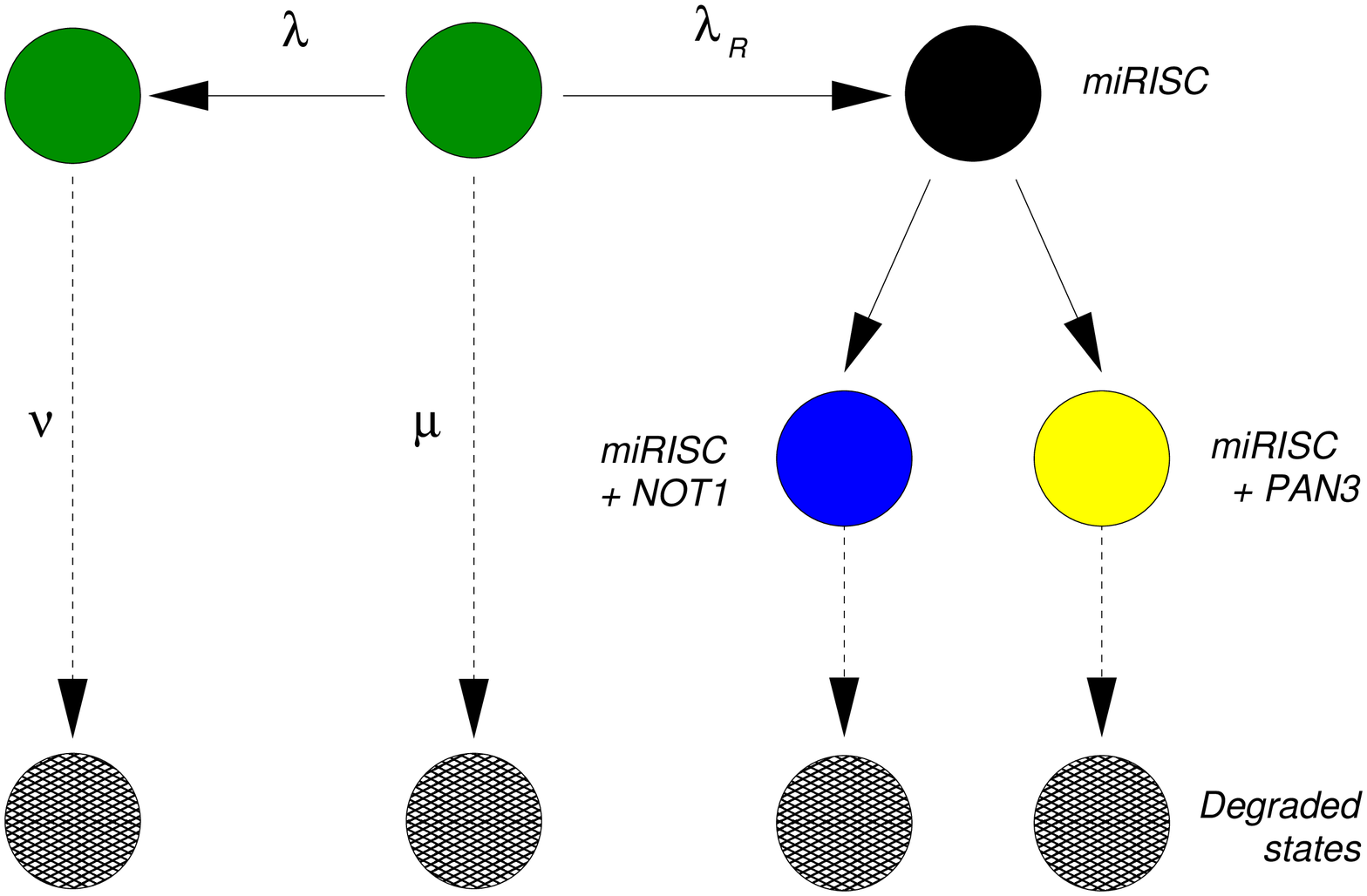} 
\end{center}
 \caption{The initial hypothesis of Braun {\em et al.} \cite{Braun2011a} is complemented with an alternative pathway that competes with the miRNA-mediated pathway. Initially, all mRNA start at the state represented by the central green circle and follow one of the two competing paths. The path towards the left (denoted by $\lambda$) is exclusive of the miRISC pathway (towards the right, denoted by $\lambda_R$), and vice versa. The rates $\lambda$, $\mu$ and $\nu$ have been fixed through the fitting performed in figure \ref{figgreen}. The negative control ``Control (-)" data result from the decay pattern of the mRNA when only the states represented by the green circles are available.}
\label{AlteNetw}
\end{figure}

\subsection{The crisis of the original hypothesis}

\begin{figure}[h!]
\begin{center}
\includegraphics[scale = 0.4]{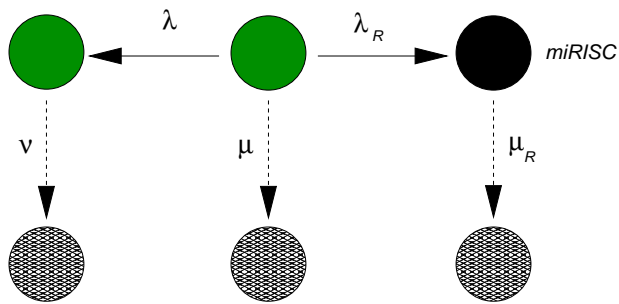}
\end{center}
\caption{After fixing the rates $\mu$, $\lambda$ and $\nu$ from the fitting of the negative decay pattern in figure \ref{figgreen}, we use this network in order to model the decay pattern when miRNA is expressed but NOT1 and PAN3 are knocked down. In order to perform this fit, however, we must add a new transition to degradation after the binding of the miRISC (rightmost transition to degradation). This new transition might contain a complex set of processes which are most likely independent of deadenylation. In the light of recent results, this arrow could include deadenylation-independent decapping \cite{Zekri2013a,Nishihara2013a}.}
\label{AlteNetwShort}
\end{figure}

At the next level of our hierarchical approach we consider the decay pattern that results when the miRNA is expressed but NOT1 and PAN3 are knocked-down (black decay pattern in figure \ref{figdata}). When PAN3 and NOT1 are knocked down, the arrows from the black state to the blue and yellow states are absent, resulting in the right pathway having no transition to degradation. However, if we look at the corresponding decay pattern (black line in figure \ref{figdata}), we realize that such a structure is not compatible with the data because a vertex without transition to degradation would imply a flattening of the curve to a steady state amount of mRNA, corresponding to the amount of mRNA arrested in this rightmost state (black circle).  To model the observed decay of mRNA, we need to postulate an additional transition from the rightmost state (represented by the black circle, after binding with miRISC) to degradation. A possible interpretation is that the additional transition (from the black circle to degradation in figure \ref{AlteNetwShort}) includes unknown biochemical degradation pathways which are independent of deadenylation. Supporting this hypothesis are the findings reported in \cite{Zekri2013a}: they report that the binding of the miRISC complex to the target mRNA can promote the dissociation of Poly-A-Binding Proteins (PABPs).  Indeed, PABPs are known to protect the poly-A tail of the mRNA from being hydrolyzed, thus stabilizing the mRNA. Thus, miRNA-mediated PABP dissociation can trigger NOT1 and PAN3-independent deadenylation, which eventually leads to the degradation of the mRNA \cite{Zekri2013a}.

\begin{figure}[h!]
\begin{center}
\includegraphics[scale = 0.6]{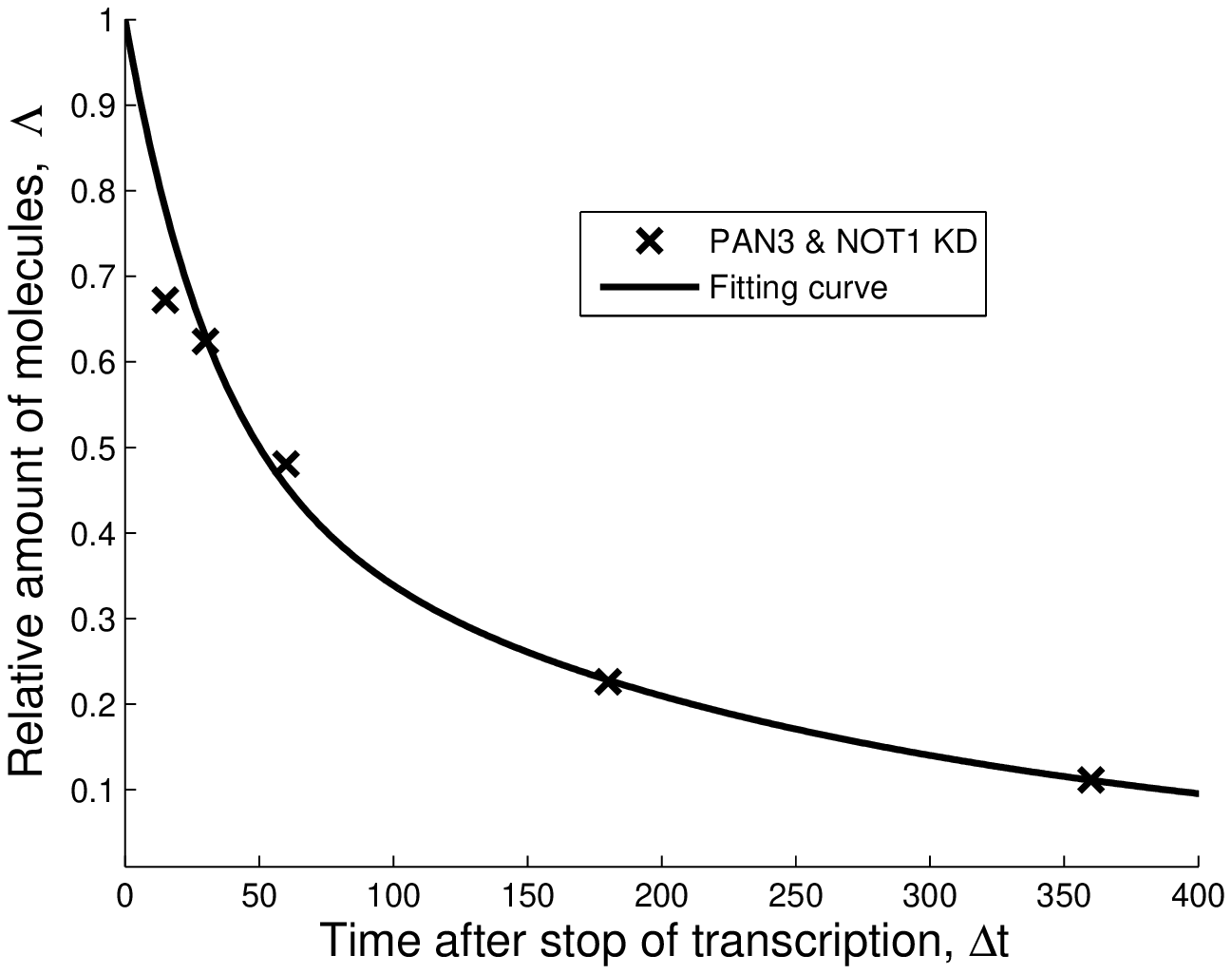} 
\end{center}
 \caption{After fixing the rates $\mu$, $\lambda$, $\nu$ from the fitting of the negative decay pattern in figure \ref{figgreen}, we use the network in figure \ref{AlteNetwShort} to fit the decay pattern when miRNA is expressed but NOT1 and PAN3 are knocked down. The values of the rates are $\lambda_R=0.0023\, \mbox{min}^{-1}$ and $\mu_R=0.0052\, \mbox{min}^{-1}$ with 95\% confidence intervals $[0,\, 0.0052]$ and $[0.0030,\, 0.0074]$, respectively. }
\label{miRISC}
\end{figure}

Fitting the data to the network depicted in figure \ref{AlteNetwShort} reveals several crucial aspects of the hypothesized network of figure \ref{fighyp} and shows the shortcomings of the latter. While the fit of the data using the network in figure \ref{AlteNetwShort} works pretty well (see figure \ref{miRISC}), it fixes the rate $\lambda_R$ associated to the binding of miRISC on the mRNA. This rate is therefore independent of the transitions occurring downstream.

The next step in our hierarchical program, however, would be to take the next decay patterns and fit them to the network given in figure \ref{AlteNetw} activating the appropriate pathway depending on which factor (NOT1 and/or PAN3) is present while keeping $\lambda_R$ fixed. Before doing that, however, a simple computation shows that the fraction of mRNA going through the miRISC pathway is given by
\beq
\sigma_R\, =\, \frac{\lambda_R}{\lambda_R+\lambda+\mu}\, \sim \, 0.074\, ,
\eeq
\ie, about 7\% of the whole mRNA binds to miRISC complexes in the absence of NOT1 and or PAN3, based on the network shown in figure \ref{AlteNetwShort}. This discovery leads to two conclusions. First, the fraction of mRNA that can be manipulated after binding with miRISC is so small that an enlargement of the network by including a separate NOT1 and a separate PAN3 pathway downstream of miRISC binding becomes meaningless. Indeed, attempts to do so lead to very poor fitting of the remaining curves (see supplementary materials). Second, such a small fraction of miRNA-regulated mRNA (about 7\%) would indicate that miRNA cannot be considered a strong mechanism of gene regulation, contrary to the experimental evidence that miRNA is a strong regulatory mechanism. Therefore, consistent with the strong role of miRNA in the regulation of mRNA, we are forced to partially reject the hypothesis formulated in figure \ref{fighyp} and revise it in search for other possible interactions between miRISC, PAN3 and NOT1. {\red Note that the computed value of 7\% is necessarily affected by some error due to the precision by which the data could be extracted from the originally published plot and by the absence of information about the biological replicas. Nevertheless, this value is an indication that the originally proposed model of degradation as described in \cite{Braun2011a} would predict that only a very small fraction of mRNA is involved in miRNA mediated degradation. In the following we will present a parsimonious model of degradation that is able to predict more realistic figures of the relative amounts of mRNAs involved in the different degradation pathways.}

Finally, the comparison between the decay pattern fitted in figure \ref{figgreen} and \ref{miRISC} shows that binding of miRISC alone does stabilize the mRNA compared to when the miRNA is not expressed. This is a strong indication that miRISC ``protects" the target mRNA from the action of alternative, competing degradation pathways.

\subsection{A new hypothesis arises from the data}

\begin{figure}[h!]
\begin{center}
\includegraphics[scale = 0.4]{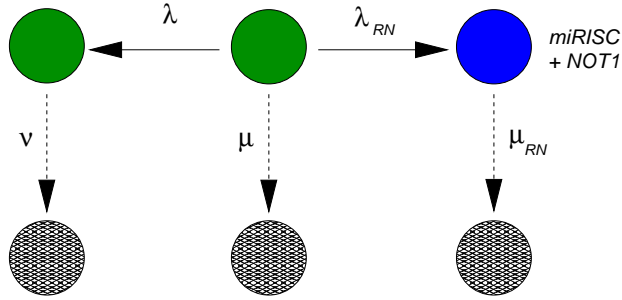} 
\end{center}
\caption{Biochemical network able to fit the data when PAN3 has been knocked down. After fixing the rates $\mu$, $\lambda$ and $\nu$ from the fitting of the negative decay pattern in figure \ref{figgreen}, we use this network in order to model the decay pattern when miRNA is expressed but PAN3 is knocked down (``PAN3 KD" data). The transition from the central green state to the state with the mRNA bound to the complex miRISC and NOT1 is ruled by the transition rate $\lambda_{RN}$. The downwards transition, ruled by the rate $\mu_{RN}$ includes several steps that cannot be specified from these data.}
\label{AlteRN}
\end{figure}

Since the initial hypothesis that miRISC binds to the mRNA and then recruits the NOT1 molecule does not result in a reasonable fit, we can hypothesize that miRISC binds to NOT1 {\em before} recruiting the target mRNA. This hypothesis is formulated in figure \ref{AlteRN}, which can be used to fit the data where only the PAN3 complex has been knocked down. 

\begin{figure}[h!]
\begin{center}
\includegraphics[scale = 0.6]{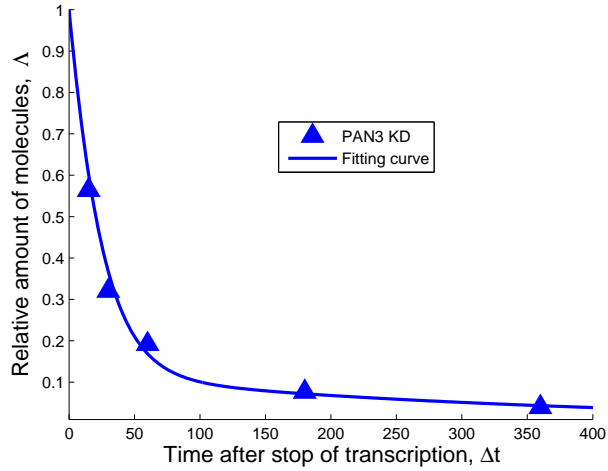} 
\end{center}
 \caption{After fixing the rates $\mu$, $\lambda$, $\nu$ from the fitting of the negative decay pattern in figure \ref{figgreen}, we use the network in figure \ref{AlteRN} in order to fit the decay pattern when miRNA is expressed but PAN3 is knocked down. The values of the rates are $\lambda_{RN}=0.046\, \mbox{min}^{-1}$ and $\mu_{RN}=0.0461\, \mbox{min}^{-1}$ with 95\% confidence intervals $[0.0305,\, 0.0687]$ and $[0.0319,\, 0.0602]$, respectively. These rates indicate that in the absence of PAN3, about 64 \% of the mRNA in the experiment are degraded by the action of miRISC.}
\label{PAN3KD}
\end{figure}

The fit is indeed very good, as seen in figure \ref{PAN3KD}.  Based on this result, the sole effect of NOT1 binding to the miRISC leads to a strong increase of the percent of mRNA that are degraded through miRISC activity, given by
\beq
\sigma_{RN}\, =\, \frac{\lambda_{RN}}{\lambda_{RN}+\lambda+\mu}\, \sim \, 0.64\, ,
\eeq
which emphasizes the strong role of NOT1 in the degradation of mRNA.

\begin{figure}[h!]
\begin{center}
\includegraphics[scale = 0.4]{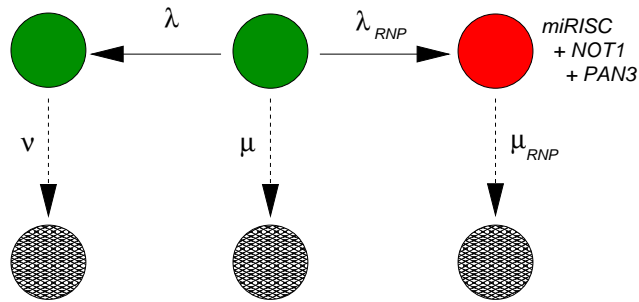} 
\end{center}
\caption{Biochemical network able to fit the data when all three factors miRISC, NOT1 and PAN3 are expressed. After fixing the rates $\mu$, $\lambda$ and $\nu$ from the fitting of the negative decay pattern in figure \ref{figgreen}, we use this network in order to model the decay pattern when all three factors are expressed. The transition from the central state (green) to the state with the mRNA bound to the complex miRISC + NOT1 + PAN3 is ruled by the transition rate $\lambda_{RNP}$. The downwards transition, ruled by the rate $\mu_{RNP}$ includes several steps that cannot be specified from these data.}
\label{AlteRNP}
\end{figure}

The final curve of the experiment in \cite{Braun2011a} concerns the action of all the factors together. On the basis of the results obtained so far in figures \ref{AlteRN} and \ref{PAN3KD} there may be several hypotheses about the possible combined action of PAN3 and NOT1. Since PAN3 alone (yellow curve in the original data shown in figure \ref{figdata}) does not have a significant effect on the decay of the mRNA compared to the action of miRISC alone, we conclude that PAN3 works cooperatively with NOT1 by forming a complex miRISC+NOT1+PAN3 before binding to the target mRNA. This hypothesis is formulated in figure \ref{AlteRNP}.

\begin{figure}[h!]
\begin{center}
\includegraphics[scale = 0.6]{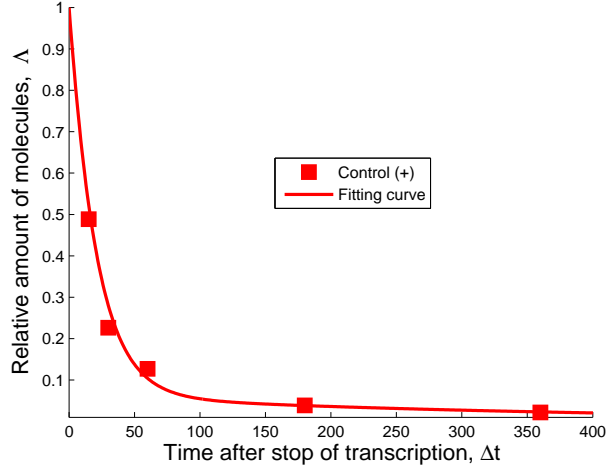} 
\end{center}
 \caption{After fixing the rates $\mu$, $\lambda$, $\nu$ from the fitting of the negative decay pattern in figure \ref{figgreen}, we use the network in figure \ref{AlteRNP} in order to fit the decay pattern when all three factors, miRNA, NOT1 and PAN3 are expressed. The values of the rates are $\lambda_{RNP}=0.1501\, \mbox{min}^{-1}$ and $\mu_{RNP}=0.0493\, \mbox{min}^{-1}$ with 95\% confidence intervals $[0.0908,\, 0.2094]$ and $[0.0373,\, 0.0612]$, respectively. These rates indicate that when both NOT1 and PAN3 are expressed together with the miRNA, about 84 \% of the mRNA in the experiment are degraded by the action of miRISC.}
\label{All}
\end{figure}

The data fits the network in figure \ref{AlteRNP} quite well, as one can see in figure \ref{All}.

By using the values $\lambda_{RNP}$ and $\mu_{RNP}$ we can again compute the fraction of target mRNA that is degraded by the action of miRISC+NOT1+PAN3:
\beq
\sigma_{RNP}\, =\, \frac{\lambda_{RNP}}{\lambda_{RNP}+\lambda+\mu}\, \sim \, 0.84\, ,
\eeq
indicating that this model produces the strong regulatory effect of the miRNA on its target as expected.

\subsection{The cooperative role of PAN3}

\begin{figure}[h!]
\begin{center}
\includegraphics[scale = 0.4]{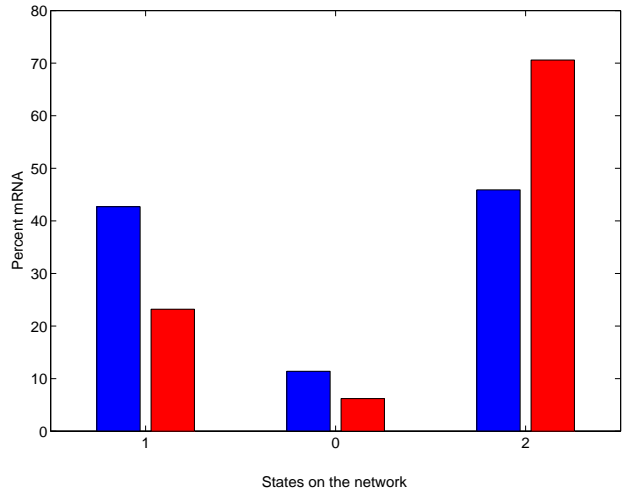} 
\end{center}
 \caption{Percent of mRNA at steady state expression level in each of the three main states of the networks in figures \ref{PAN3KD} and \ref{All}. State $0$ represents the mRNAs that are not bound to miRISC and not bound to the competing alternative complexes. The alternative degradation pathways lead to state $1$ whereas the miRISC-mediated pathway leads to state $2$. The blue histogram refers to the case when only the miRNA and NOT1 are expressed whereas the histogram in red refers to the experiments when all three factors miRNA, NOT1 and PAN3 are expressed. The percent of mRNA in state $2$ is strongly increased by the expression of PAN3 at the expenses of the amount of mRNA in state $1$ This indicates that the most important role of PAN3 is to shift the balance towards miRNA-mediated decay.}
\label{histo}
\end{figure}

We have seen that the expression of PAN3 in a system with miRNA and NOT1 strongly destabilizes the target mRNA and shortens its lifetime. Nevertheless, we can better understand the role played by PAN3 in cooperation with NOT1, when we compare the fraction of target mRNA that are expected to be found in the "miRISC+NOT1" state in figure \ref{PAN3KD} (blue circle) with the fraction of target mRNA to be found in the "miRISC+NOT1+PAN3" state in figure \ref{All} (red circle). This comparison is made in figure \ref{histo}. There, we find the fraction of mRNAs in each of the three states after denoting state $0$ the state in the middle of the network, state $1$ the state on its left side and state $2$ the state on the right side (blue circle in figure \ref{PAN3KD} and red circle in figure \ref{All}). 

We can see from the bar plot that the major contribution of PAN3 is to shift the balance of forces in favor of the miRNA by subtracting target mRNAs to the alternative pathway. By expressing PAN3, indeed, the amount of mRNA that are found in state $1$, corresponding to the mRNA bound to protein complexes competing with the miRISC, decreases by almost 20\% of the total mRNA, whereas the amount found in state $0$ decreases by only a 5\% of the total. This might indicate that the major role played by PAN3 is not to enhance deadenylation but rather to enhance the recruitment of the target mRNA at the expenses of alternative degradation pathways that do not involve miRNA. From the available data it is not possible to establish if the mRNA in these three states are also translational competent or are silenced. From the biochemical point of view, moreover, each of these three states might be a complex of different states sharing the same kinetic characteristics. Nevertheless, experiments designed to estimate the amount of mRNA bound or not bound to miRISC and NOT1 can provide important information to validate this model.

\section{Summary and Discussion}
In this paper, we show that the current hypothesis about the sequence of interactions between miRISC, its target mRNA and the factor NOT1 is not supported by the data. We have shown that the mRNA is also degraded when the miRNA is not expressed, indicating the existence of an alternative pathway, possibly competing with the miRNA pathway. 

We also show that when only miRNA is expressed (NOT1 and PAN3 are knocked down), the target mRNA is stabilized, probably because it is protected from the action of an alternative miRNA-independent pathway. We postulate that the binding between miRISC and mRNA is irreversible and leads to the deadenylation independent decay of the target message in agreement with recent experimental studies. However, this assumption is not obligatory. Indeed, one could have hypothesized that binding to miRISC is reversible, and that the presence of miRNA alone just slows down the action of the alternative pathway. With the present data it is not possible to distinguish between these two alternatives.

Finally, our analysis indicates that the miRISC complex and NOT1 interact with each other {\em before} interacting with the mRNA.  We assume that this discovery is not limited to the special miRNA-mRNA pair studied in \cite{Braun2011a} and is therefore a new general mechanism of mRNA control. Our analyses confirm the conclusions in \cite{Braun2011a} that PAN3 without NOT1 does not lead to an identifiable destabilization of the mRNA. Nevertheless, we see a strong cooperative effect between PAN3 and NOT1, where PAN3 is able to strongly enhance the binding of the miRISC+NOT1+PAN3 complex to the target mRNA compared to the miRISC+NOT1 complex alone.

Experimentally, one should be able to detect the presence of miRISC+NOT1 complexes in the absence of target mRNA, in order to verify our findings. Moreover, steady state relative amounts of mRNA in the different biochemical states can provide further validation data for our networks and additional information to unveil further details of the miRNA-mediated mRNA degradation.
\section*{Acknowledgments}
The authors acknowledge support from the ITN Marie-Curie ``NICHE".
\appendix
\section{Supplementary Materials}
\subsection{From single-molecule pathways to decay patterns}
We model the mRNA decay as a single molecule stochastic process. The process starts from an initial state $\sigma_0$, which represents the mRNA in its initial condition, and terminates in the degradation state $\sigma_n$. The initial and the degradation states are defined here based on the time point from which the mRNA is experimentally detected until it is not detected anymore, respectively, in the context of the experimental technique used in \cite{Braun2011a}. 
Each realization of the stochastic process, which starts from $\sigma_0$ to the degradation state through the network of states represents the life of a single mRNA molecule.
Therefore, the distribution of the random time $T$ elapsing from $\sigma_0$ to the degradation state $\sigma_n$ can be identified with the distribution of the lifetimes of the mRNA molecules. The state spaace $\sigma$ is thus made of $n$ transient states and $1$ absorbing state $\sigma_n$ (a generalization to more absorbing states is strightforward, see \cite{Valleriani2014a}).

In our modeling framework each state transition describes the occurrence of what we define a {\em first-order biochemical event}. This definition includes three possible biochemical scenarios:
\begin{enumerate}[(i)]
\item an elementary first-order chemical reaction
\item a pseudo-first order biochemical reaction: the actual reaction order is $>1$ but the concentration all of the reactants except for one are at the steady state level. As a result, this reaction behaves as a ``true'' first order reaction
\item an apparent first-order reaction: a complex chain of reactions with a single rate limiting step, thus exhibiting first-order behavior
\end{enumerate}

In the context of the mRNA degradation pathway we consider the two latter cases only, due to the complexity of the involved biochemical reactions.
If we assume that the probability of occurrence of each biochemical event depends only on the current state (Markov property), then we can model the stochastic processes describing the life of mRNA molecules as Markov chains.          

Thus, following \cite{Valleriani2014a, Valleriani2008a, Keller2012a, Deneke2013a}, we can write the lifetime probability density $\phi$ as
\beq
\phi(t)\, =\, \vec{e}_0 \left[ \exp \left( \mathbf{S}_0 t \right) S_A\right]\, ,
\label{phi}
\eeq
where $S_0$ is a square $n\times n$ matrix. The off-diagonal elements of $S_0$ are the transition rates between the transient states, and the diagonal elements are composed of the negative sum over all outgoing transition rates from each transient state. The column vector $S_A$ contains all the transition rates regarding transitions from the transient into the absorbing state. The row vector of length $n$ $\vec{e}_0$ has a one in the position corresponding to the initial condition and zeros elsewhere.

\begin{figure}[h!]
\begin{center}
\includegraphics[scale=0.7]{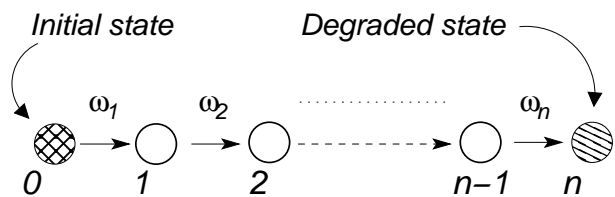} 
\end{center}
\caption{A generic chain made of $n$ irreversible transitions. The probability density of the random time $T$ from the initial to the degraded state is given by $\phi_n(t)$ derived in Eq.\ (\ref{phin}). }
\label{S1}
\end{figure}

Similar to \cite{Deneke2013a}, we must take into consideration that at the beginning of the experiment, the expression of mRNA is at the steady state. Thus, not all mRNA molecules are at the initial state; some of them will be in other states because they are older. If the transcription of mRNAs was ongoing long before the beginning of the decay experiment, one can assume that the age distribution of the population of mRNA is at steady state\footnote{Short transcription pulses instead generate an age distribution away from steady state. This interesting extension of the theory will not be considered here because it is not relevant in this context.}. In \cite{Deneke2013a} it was shown that when the age distribution is at steady state, the fraction of mRNA remaining $\Delta t$ time units after the stop of transcription is given by the function 
\beq
\label{Lambda}
\Lambda(\Delta t)\, =\, \frac{1}{\mbox{E}(T)}\int_{\Delta t}^\infty \de t\, \left( 1-\Phi(t)\right)\, ,
\eeq
where $\mbox{E}(T)$ is the average lifetime, namely the average value of $T$ given by
\beq
\mbox{E}(T)\, =\,\int_0^\infty \de t\, \left( 1-\Phi(t)\right)\, ,
\eeq 
and $\Phi$ is the probability function of $T$ given by
\beq
\Phi(t)\, =\, \int_0^t\de\tau \, \phi(\tau)\, .
\eeq
An important technical point for the computation of $\Lambda$ is that in Eq.\ (\ref{Lambda}), $\mbox{E}(T)$ is just the normalization factor in order to have $\Lambda(0)=1$.
The networks used in the main article are sufficiently simple, so that the calculation of $\phi$ can be done easily (see below), without recurring to the relatively complex formulation given in Eq.\ (\ref{phi}). Nevertheless, Eq.\ (\ref{phi}) is the general form of the function, which depends on the values of the rates in the matrix $S_0$. When Eq.\ (\ref{Lambda}) has been written in terms of the rates, the rates can be determined by means of a nonlinear fitting of the log of $\Lambda(\Delta t)$ with the log of the data. The use of the logged data is justified by the common assumption that the experimental noise is multiplicative.  Eqs.\ (\ref{phi}) and (\ref{Lambda}) include the case when there is just one transient state. This is the case when the decay pattern follows an exponential function whose parameter is the rate of degradation, and also the inverse of the average lifetime. 

\subsubsection{Fitting functions: General aspects}

All our fits were performed for two or three parameters and compared, where it made sense, with the exponential fit. Based on the Akaike Information Criterion (AIC) corrected for small sample sizes, we found that the fits with the exponential function did not perform sufficiently well to be considered in this study.

Since all our networks are comprised of irreversible transitions only, the computation of $\Lambda$ can be split in the sub-tasks of computing the contribution of chains of $1,2,\ldots,n$ transitions such as the one in figure \ref{S1}. For each of these chains, the probability density $\phi_n$ for the absorption time in state $n$ can be computed as the convolution of $n$ independent exponential functions, and it is given as
\beq
\label{phin}
\phi_n (t) \, =\, \left(\prod_{k=1}^n \omega_k\right)\cdot\left( \sum_{j=1}^n \frac{\exp(-\omega_j t)}{\prod_{k\neq j} (\omega_k - \omega_j)}\right)\, ,
\eeq
if all the $\omega_j$ are different from one another. In the networks considered in this work we just need the two cases corresponding to $n=1$, \ie the exponential function, and $n=2$. The cumulative functions associated to the $\phi_n$ are thus given by
\beq
\label{Phin}
\Phi_n(t)\, =\, 1\, -\, \sum_{j=1}^n \left(\prod_{k\neq j}\frac{ \omega_k}{\omega_k - \omega_j}\right) \exp(-\omega_j t)\, ,
\eeq
again under the assumption that all rates are different. As we have seen after fitting the data, this assumption is never violated.

Assume now that a given network has, say, three possible paths from the initial state to the absorbing state made of only irreversible transitions. Let us call these three paths $a$, $b$, and $c$ and let us assume that these three paths are characterized each by its own set of $\omega$'s and that the probability of each path to be taken is $p_a$, $p_b$ and $p_c$, respectively. Let $\phi_a$, $\phi_b$ and $\phi_c$ be the probability densities conditioned on each of the paths separately. Then, the total probability density $\phi$ is given by
\beq
\phi\, =\, p_a\phi_a+p_b\phi_b+p_c\phi_c\, ,
\eeq
and the cumulative probability function is given by
\beq
\Phi\, =\, p_a\Phi_a+p_b\Phi_b+p_c\Phi_c\, ,
\eeq
so the decay function $\Lambda$ from Eq.\ (\ref{Lambda}) reads
\beq
\Lambda(\Delta t) \, \propto\,  \int_{\Delta t}^\infty \de t\, \left(1-\sum_x p_x \Phi_x(t)\right)\, ,
\eeq
where $x=a,b,c$, and the proportionality constant must be fixed so that $\Lambda(0)=1$. If we now write the $\Phi_x$ as $\Phi_x+1-1$ and we substitute, we obtain
\beq
\label{Lambdax}
\Lambda(\Delta t)\, \propto \, \sum_x p_x \mathcal{O}_x(\Delta t;\vec{\omega}_x)\, ,
\eeq
where we have defined
\beq
\label{Ox}
\mathcal{O}_x(\Delta t; \vec{\omega}_x)\, =\, \int_{\Delta t}^\infty \de t\, \left(1-\Phi_x(t)\right)\, .
\eeq
The formulation in Eq.\ (\ref{Lambdax}) turns out to be a quite useful, since our $\Phi_n$ expressed in Eq.\ (\ref{Phin}) take an explicit relatively simple form once put in Eq.\ (\ref{Ox}). The vector $\vec{\omega}_x$ is the set of all $\omega$'s along the path $x$. Upon simple explicit integration we have thus, for $n=1$ and $n=2$ the two functions
\beq
\mathcal{O}_1(\Delta t; \omega_1) = \frac{\exp (-\omega_1 \Delta t)}{\omega_1}\, ,
\eeq
and 
\beq
\mathcal{O}_2(\Delta t; \omega_1, \omega_2) = \frac{\omega_2\exp (-\omega_1 \Delta t)}{\omega_1 (\omega_2 - \omega_1)}+\frac{\omega_1\exp (-\omega_2 \Delta t)}{\omega_2 (\omega_1 - \omega_2)}\, ,
\eeq
to be used in the explicit calculations that follow.

\subsubsection{Fitting functions: Specific forms}
This section delivers the explicit form of the functions used in the fitting of the data.

\begin{itemize}
\item The generic two-state network in Figure 3 in the main article has
\beq
\label{Laneg}
\begin{array}{lcl}
\bds \Lambda(\Delta t)\eds & \propto & \bds\frac{\mu}{\mu+\lambda}\mathcal{O}_1(\Delta t;\omega_1=\lambda+\mu)\eds\\
& & \\
&+&\bds \frac{\lambda}{\mu+\lambda}\mathcal{O}_2(\Delta t; \omega_1=\lambda+\mu, \omega_2 = \nu)\eds\, ,
\end{array}
\eeq
where in both cases $\omega_1=\lambda+\mu$ takes into account that the dwell time on state $A$ is exponential with parameter $\lambda+\mu$. This function was used to fit the data of the negative control (green line) and therefore to fix the rates $\lambda$, $\mu$ and $\nu$. These three rates are kept fixed in all other networks.

\item The networks in Figures 6, 8, 10 in the main article share the same structure and lead to 
\beq
\label{Lafac}
\begin{array}{lcl}
\bds\Lambda(\Delta t)\eds & \propto & 
\bds\frac{\lambda}{\lambda+\mu+\lambda^\prime}\mathcal{O}_2(\Delta t; \omega_1 = \lambda+\mu+\lambda^\prime, \omega_2 = \nu)\eds \\
& & \\
& + &\bds \frac{\mu}{\lambda+\mu+\lambda^\prime}\mathcal{O}_1(\Delta t; \omega_1 = \lambda+\mu+\lambda^\prime)\eds \\
& & \\
& + &\bds \frac{\lambda^\prime}{\lambda+\mu+\lambda^\prime}\mathcal{O}_2(\Delta t; \omega_1 = \lambda+\mu+\lambda^\prime, \omega_2 = \mu^\prime)\eds\, ,
\end{array}
\eeq
where $\lambda^\prime = \lambda_R, \,\lambda_{RN},\, \lambda_{RNP}$ and $\mu^\prime=\mu_R, \,\mu_{RN},\, \mu_{RNP}$ in Figures 6, 8, 10, respectively. To fit the data, as was done in Figures 7, 9, and 11 in the main article, we have kept the three rates $\lambda$, $\mu$ and $\nu$ fixed as they have been determined in the fit of the negative control. This means that the $\Lambda(\Delta t)$ given in Eq.\ (\ref{Lafac}) was used to fit the two additional rates $\lambda^\prime$ and $\mu^\prime$ for each of the following three data sets.
\end{itemize}
\begin{table}
\begin{center}
  \begin{tabular}{| r || c | c | c | c | c |}
    \hline
$\Delta t$ & (-) & R & RP & RN & RNP \\ \hline \hline 
    0 & 1 & 1 & 1 & 1 & 1\\ \hline
    15 & 0.6899 & 0.6720 & 0.6722 & 0.5641 & 0.4885\\ \hline
    30 & 0.5588 & 0.6238 & 0.6322 & 0.3206 & 0.2263\\ \hline
    60 & 0.3375 & 0.4806 & 0.4184 & 0.1919 & 0.1272\\ \hline
    180 & 0.1462 & 0.2262 & 0.2182 & 0.0766 & 0.0388\\ \hline
    360 & 0.0857 & 0.1112 & 0.1035 & 0.0397 & 0.0217\\ \hline
  \end{tabular}
\end{center}
\caption{\label{table1} Experimental data as extracted from the plot published in \cite{Braun2011a}. The various experimental conditions are given in the first row: (-) = negative control, when the miRNA is not expressed; R = when only miRNA is expressed but both NOT1 and PAN3 are not expressed; RP = data when NOT1 is knocked down; RN = data when PAN3 is knocked down; RNP = all factors are present, no knock downs.}
\end{table}
\subsubsection{Data used in the study}

The experimental data used in this study have been extracted from \cite{Braun2011a}. We report the numerical values in table \ref{table1}:
the first column is the measurement time expressed in minutes; the second column is the data for the negative control, when the miRNA is not expressed, which was fitted with Eq.\ (\ref{Laneg}); the third column is the data when both NOT1 and PAN3 are knocked down (only miRNA is expressed) and was fitted with Eq.\ (\ref{Lafac}); the fourth column is the data when NOT1 is knocked down: since this data is very close to the data in the third column, an additional fit of it was not performed as these two sets of data cannot be distinguished from each other; the fifth column is the data when PAN3 is knocked down and was fitted with Eq.\ (\ref{Lafac}); the last column is the positive control, when all factors are expressed and was fitted with Eq.\ (\ref{Lafac}).
\begin{table}
\begin{center}
  \begin{tabular}{| c || r | c |}
    \hline
rate & $10^{-3} \mbox{min}^{-1}$ & $10^{-3} \mbox{min}^{-1}\, 95\%$ CI \\ \hline \hline 
    $\lambda$ & 0.8 & [0.2, 1.3] \\ \hline
    $\mu$ & 27.6 & [22.9, 32.4] \\ \hline
    $\nu$ & 2.8 & [1.8, 3.8] \\ \hline
    $\lambda_R$ & 2.3 & [0.0, 5.2] \\ \hline
    $\mu_R$ & 5.2 & [3.0, 7.4]\\ \hline
    $\lambda_{RN}$ & 46.0 & [30.5, 68.7] \\ \hline
    $\mu_{RN}$ & 46.1 & [31.9, 60.2] \\ \hline
    $\lambda_{RNP}$ & 150.1 & [90.8, 209.4] \\ \hline
    $\mu_{RNP}$ & 49.3 & [37.3, 61.2] \\ \hline
  \end{tabular}
\end{center}
\caption{\label{table2} Values of the rates as they arise from the fitting of the experimental data. By using our derived functions $\Lambda$, Eqs.\ (\ref{Laneg}) or (\ref{Lafac}), we can perform a non-linear fit of the data listed in table \ref{table2}. The values of the rates are given together with their 95\% confidence interval (CI). }
\end{table}
\subsubsection{Fitting procedure and algorithm}

Our fitting procedure consisted in finding the set of parameters $\vec{\omega}$ that minimized the square deviation of the logarithm of the data, as
\beq
\mbox{RSS} \, =\, \sum_{\{\Delta t\}} \left(\log (\mbox{data} (\Delta t) ) - \log (\Lambda (\Delta t))\right)^2\, ,
\eeq
where ``data" is any of the columns in the table above and $\Lambda$ is either Eq.\ (\ref{Laneg}) or Eq.\ (\ref{Lafac}) as explained above. The choice of the logarithm is dictated by the assumption that the noise is multiplicative. The minimization was done by using standard MATLAB routines (``lsqnonlin" and ``nlinfit") leading to the same results. By using ``nlinfit" we were able to estimate also the 95\% confidence interval, reported together with the estimated values in the table \ref{table2}:

Since the data concerns averages whereas the biochemical model for the degradation pathway is based on a single molecule perspective, it is to be expected that the confidence intervals are relatively large. In addition, the data used had to be extracted graphically from the plots published in \cite{Braun2011a} and thus are expected to contain random errors related to reading the data from the plots. Despite these shortcomings, any other of the alternative models we have tested were not able to fit the data (not shown), thus indicating that the networks proposed in this study are structurally correct.

\begin{table}
\begin{center}
  \begin{tabular}{| c || r | r | r | r | r |}
    \hline
condition & $\mbox{E}[T]$ min & $ \%$ via miRISC & $100\pi_0$ & $100\pi_1$ & $100\pi_2$\\ \hline \hline 
    (-) & 44.7 & 0.0 & 78.8  & 21.2  & 0 \\ \hline
    R & 55.6 &  7.4 & 58.6 & 15.7 &  25.6 \\ \hline
    RP & 52.7 & 5.5 & 63.1 & 16.9  &19.9\\ \hline
    RN & 30.1 & 63.6 & 42.7 & 11.4 & 45.9\\ \hline
    RNP & 24.2 &  84.1 & 23.2 & 6.2 & 70.6\\ \hline
  \end{tabular}
\end{center}
\caption{\label{table3} Steady state properties. For each condition we compute the average lifetime of the mRNA (second column), the percent of molecules that are degraded via miRISC (third column) and the distribution of mRNA among the three main biochemical states of figure \ref{S2}.}
\end{table}

\subsubsection{Steady state properties and statistics}

We can associate to each path of the type drawn in figure \ref{S1} its average absorption time given by
\beq
\mbox{E}[T_n]\, =\, \sum_{j=1}^n \omega_j^{-1}\, ,
\eeq
which allows us to compute the average lifetime associated to each of the networks discussed in our paper. Therefore, the average lifetime associated to the decay pattern described by Eq.\ (\ref{Laneg}) is given by
\beq
\mbox{E}[T^{(-)}]\, =\, \frac{\nu+\lambda}{\nu(\lambda+\mu)}\, ,
\eeq
where the upper index (-) indicates that this average lifetime refers only to the negative control decay data. Conversely, the average lifetime associated to the decay patterns described by Eq.\ (\ref{Lafac}) is given by
\beq
\mbox{E}[T^{(RX)}]\, =\, \frac{\lambda\mu^\prime+\lambda^\prime \nu+\mu^\prime \nu}{\mu^\prime\nu(\lambda+\mu+\lambda^\prime)}\, ,
\eeq
where the upper index (RX) indicates that this refers to all analyzed networks where miRNA is involved together with none, one or both factors PAN3 and NOT1.
\begin{figure}[h!]
\begin{center}
\includegraphics[scale=0.5]{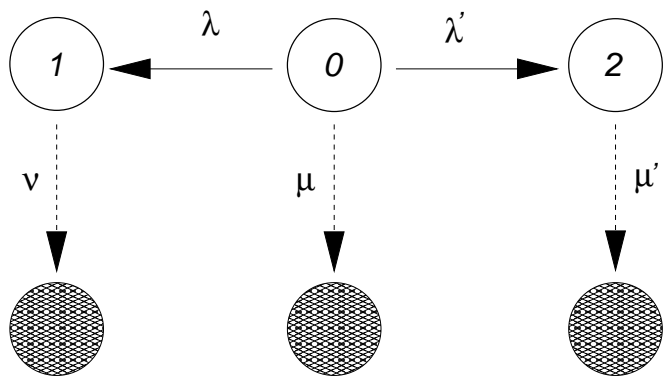} 
\end{center}
\caption{The network studied in this contribution is made of three states, $0$, $1$ and $2$, whereas state $2$ cannot be reached when $\lambda^\prime = 0$ as in the case of the negative control. Using the rates and the Master equation we can compute the stationary distribution of the mRNA's in the three states for all models considered. }
\label{S2}
\end{figure}
When the rates are known one can compute the percent of target mRNA that are involved in miRNA mediated degradation, shown in the third column of table \ref{table3}. Notice that the halftimes computed in \cite{Braun2011a} are very similar in value to our average lifetimes. The average lifetimes, however, have a clear meaning here because the steady state amount of mRNA is proportional to the average lifetime, whereas the halftimes only have a meaning in the framework of the exponential decay.

Another quantity of biological interest is the percent of mRNA in the different biochemical states of our network at steady state. The three states of the network are drawn in figure \ref{S2} and the stationary probabilities $\pi_0$, $\pi_1$ and $\pi_2$ can be computed from the Master equation by redirecting the arrows with rates $\mu$, $\nu$ and $\mu^\prime$ towards the initial state $0$. This leads to the stationary distribution
\beq
\begin{array}{lcl}
\pi_0 & = & \nu\mu^\prime / (\nu\mu^\prime + \lambda\mu^\prime + \lambda^\prime \nu) \\
\pi_1 & = & \lambda\mu^\prime / (\nu\mu^\prime + \lambda\mu^\prime + \lambda^\prime \nu) \\
\pi_2 & = & \nu\lambda^\prime / (\nu\mu^\prime + \lambda\mu^\prime + \lambda^\prime \nu) 
\end{array}
\eeq
which have a clear limit when $\lambda^\prime\to 0$ for the negative control case. The values of the probability distribution can be used to estimate how many mRNA should be found in the cell in the three different biochemical states, $0$, $1$ and $2$ in figure \ref{S2}. Notice that the \% of mRNA degraded through miRNA targeting can also be computed in terms of normalized fluxes:
\beq
\mbox{\% via miRISC}\, =\, \frac{\pi_2\mu^\prime}{\pi_0\mu+\pi_1 \nu + \pi_2\mu^\prime}\, ,
\eeq
consistent with the values computed earlier, given in the third column of table \ref{table3}
\begin{figure}[h!]
\begin{center}
\includegraphics[scale=0.5]{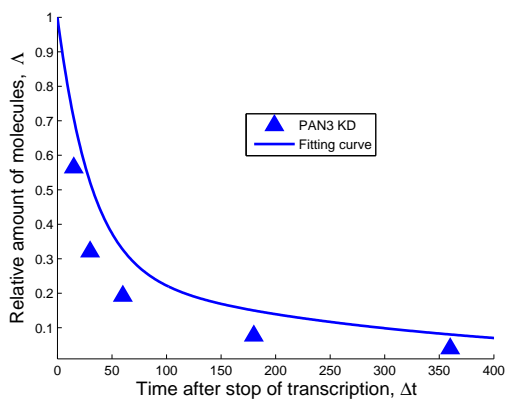} 
\end{center}
\caption{Fit of the data of the decay pattern when PAN3 has been knocked down. The model used here is the one of \cite{Braun2011a} complemented with a pathways for alternative routes to degradation. In our interpretation this means that $\lambda^\prime = \lambda_R$ in figure \ref{S2} is fixed with the data of the experiment with both PAN3 and NOT1 knocked down, while also $\lambda$, $\mu$ and $\nu$ are kept fixed. The only parameter left free to be fixed with this set of data is therefore $\mu^\prime$. However, since too few mRNAs are targeted by miRISC alone, the extra parameter $\mu^\prime$ is not sufficient to fit the data.}
\label{S3}
\end{figure}
\subsection{Why the originally hypothesized pathway is not consistent with the experimental data}
\begin{figure}[h!]
\begin{center}
\includegraphics[scale=0.5]{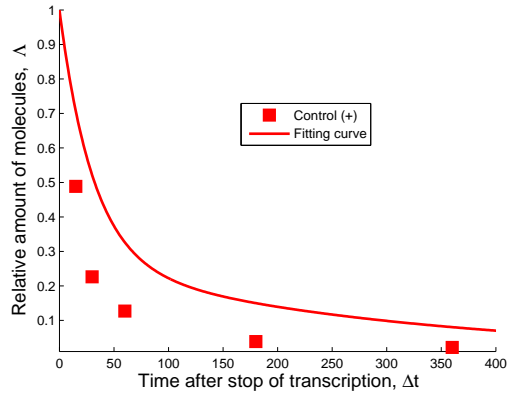} 
\end{center}
\caption{Fit of the data of the decay pattern when all factors are expressed (no knock downs). The model used here is the one proposed in \cite{Braun2011a} complemented with a pathway for alternative routes to degradation. In our interpretation this means that $\lambda^\prime = \lambda_R$ in figure \ref{S2} is fixed with the data of the experiment with both PAN3 and NOT1 knocked down, while also $\lambda$, $\mu$ and $\nu$ are kept fixed. The only parameter left free to be fixed with this new set of data is therefore $\mu^\prime$. However, since too few mRNAs are targeted by miRISC alone, the extra parameter $\mu^\prime$ turns out to be not sufficient to fit the data.}
\label{S4}
\end{figure}

 There are several reasons why the biochemical network proposed in \cite{Braun2011a} is not an adequate model for miRNA-mediated mRNA degradation: 
\begin{enumerate}
\item The network proposed in \cite{Braun2011a} does not contain all the state transitions that are necessary for fitting both the negative control data and the data where the miRNA is knocked-down.
\item Through the action of miRISC alone, only about 7\% of the target mRNA are degraded via the action of the miRNA (see second row, third column  in table \ref{table3})
\item The hypothesis formulated by \cite{Braun2011a}  foresees that NOT1 and PAN3 bind to the mRNA after the binding of miRISC. To mimic this scenario, we can thus impose $\lambda^\prime = \lambda_R$ in figure \ref{S2}, thereby implying that only one degree of freedom, namely the parameter $\mu^\prime$, remains available for being tuned through the data fitting procedure.  As seen in figures \ref{S3} and \ref{S4}, the model proposed with this assumption does not provide a good fit to the data.
\end{enumerate}

\bibliographystyle{bmc-mathphys} 
\bibliography{biblio}
\end{document}